\documentclass[aps,twocolumn]{revtex4}
\usepackage{latexsym}
\usepackage{amsfonts}
\usepackage{graphicx} 
\usepackage{epsfig}
\begin{document}

\title{Relation between the Liquid-Liquid Phase Transition\\ and Dynamic
Behavior in the Jagla Model}

\author{
Limei Xu$^1$, Isaac Ehrenberg$^2$,
Sergey V. Buldyrev$^{2,1}$, H. Eugene Stanley$^1$  
}

\bigskip

\affiliation{$^1$ Center for Polymer Studies and Department of Physics,
Boston University, Boston, MA 02215 USA\\ $^2$ Department of Physics, Yeshiva
University, 500 West 185th Street, New York, NY 10033 USA \\ }


\date{\today ~~ xebs.tex}


\begin{abstract}

Using molecular dynamics simulations, we study a spherically-symmetric
``two-scale'' Jagla potential with both repulsive and attractive ramps. This
potential displays a liquid-liquid phase transition with a positively sloped
coexistence line ending at critical point well above the equilibrium melting
line. We study the dynamic behavior in the vicinity of this liquid-liquid
critical point.  We find that the dynamics in the more ordered high density
phase (HDL) are much slower then the dynamics in the less ordered low density
phase (LDL). Moreover, the behavior of the diffusion constant and relaxation
time in the HDL phase follows approximately an Arrhenius law, while in the
LDL phase the slope of the Arrhenius fit increases upon cooling. On the other
hand, if we cool the system at constant pressure above the critical pressure
behavior of the dynamics smoothly changes with temperature. It resembles
the behavior of the LDL at high temperatures and resembles the behavior of
the HDL at low temperatures. This dynamic crossover happens in the vicinity
of the Widom line (the extension of the coexistence line into the one-phase
region) which also has a positive slope. Our work suggests a possible general
relation between a liquid-liquid phase transition and the change in dynamics.

\end{abstract}

\maketitle
\newpage
  
\section{Introduction} 

\noindent
Water is the most important solvent for biological function
\cite{Bellisent,Robinson1996}, yet it possesses many properties that are not
well understood. An open question of general interest concerning liquid water
is the relation between a liquid-liquid (LL) phase transition
\cite{poole1,poole2} and dynamic properties \cite{chenJCP2004, chen2005PC,
PNAS,Mallamace_PRL_2005, ChenPNAS, Pradeepnature}. Based on analogies with
other network forming liquids and with the thermodynamic properties of the
amorphous forms of water, it has been suggested that, at ambient pressure,
liquid water should show a crossover between fragile behavior at high T to
strong behavior at low T
\cite{Bergman00,Angell93,Itonature1999,chenJCP2004,chen2005PC,Mallamace_PRL_2005}
in the deep supercooled region of the phase diagram below the homogeneous
nucleation line. This region, called the ``no-man's land,'' may contain a LL
critical point \cite{poole1}, the terminal point of a line of first order LL
phase transitions. One current hypothesis concerns the possibility that
water's anomalies are related to the existence of this LL critical point and
its associated LL phase transition
line~\cite{poole1,PNAS,Sciortino2003}. Thus far there has been no direct
experimental proof of this hypothesis, but recent experiments in nanoscale
hydrophilic pores \cite{chen2005PC} showing a line of sharp crossover in
water diffusivity were interpreted in terms of the effects of the LL critical
point. A dynamic crossover has also been associated with the LL phase
transition in silicon and silica \cite{Poolenature2001,sastrynature2003}. The
relation between LL phase transition and dynamics properties may be not
limited to tetrahedral liquids, but is a general feature of all liquids near
the LL critical point. Here we review the results of Ref.~\cite{PNAS} on the
Jagla model of liquid which models intermolecular interactions using a
spherically-symmetric soft-core potential. We also study phase segregation
and dynamics below the LL critical point.


\section{Two-scale spherically-symmetric Jagla ramp potential}

\begin{figure}[htb]
\includegraphics[width=6cm]{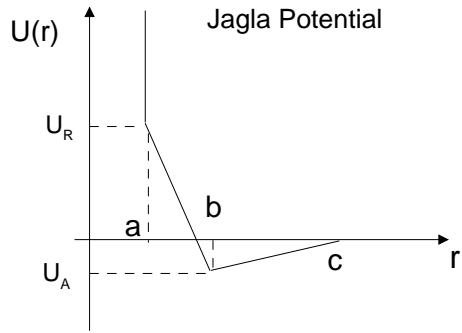}
\caption{The ``two-scale'' Jagla ramp potential with attractive and repulsive
ramps. Here $U_{R}=3.5U_{0}$, $U_{A}=-U_{0}$, $a$ is the hard core diameter,
$b=1.72a$ is the soft core diameter, and $c=3a$ is the long distance
cutoff. In the simulation, we use $a$ as the unit of length, and $U_{0}$ as
the unit of energy.}
\label{fig:jagla-pot}
\end{figure}

A spherically-symmetric potential with two different length scales has been
studied recently \cite{Sadr98,Scala01,Buldyrev02,
buldyrev02,Stell72,Jagla99,franzese2001nature, Wilding2002,Wilding2006,yan_PRL_2005,PNAS}. Here, we
study the linear ramp potential with both attractive and repulsive parts
\cite{Jagla99}. The potential is defined as

\begin{equation}
U(r) = \left\{
\begin{array}{ll}
\infty & r < a\\ U_A+(U_A-U_R)(r-b)/(b-a) & a<r<b,\\ U_A(c-r )/(c-b) &
b<r<c,\\ 0 & r > c
\end{array}\right.
\label{eq:potential}
\end{equation}

where $U_{R}=3.5U_{0}$ is the repulsive energy, $U_{A}=-U_{0}$ is the
attractive part, $a$ is the hardcore diameter, $b=1.72a$ is the well minimum,
and $c=3a$ is the cutoff at large distance [Fig.~\ref{fig:jagla-pot}].

We approximate the potential of Eq.(~\ref {eq:potential})
by a step function with small steps $\Delta U=U_0/8$ and
implement the standard discrete molecular dynamics algorithm for particles
interacting with step potentials~\cite{rapaport,Buldyrev2003physicA,
franzese2001nature}, We use $a$ as the unit of length, particle mass $m$ as
the unit of mass and $U_{0}$ as the unit of energy. The simulation time is
therefore measured in units of $a\sqrt{m/U_{0}}$, temperature in unit of
$U_{0}/k_{B}$, pressure in units of $U_{0}/a^{3}$ and density $\rho$ in units
of $Na^{3}/L^{3}$, where $L$ is the size of the system and $N=1728$ is the
number of particles. We use the Berendsen thermostat and barostat for
constant temperature and constant pressure simulations
\cite{berendsenJCP1984}.


\begin{figure}[htb]
\includegraphics[width=8cm,angle=0]{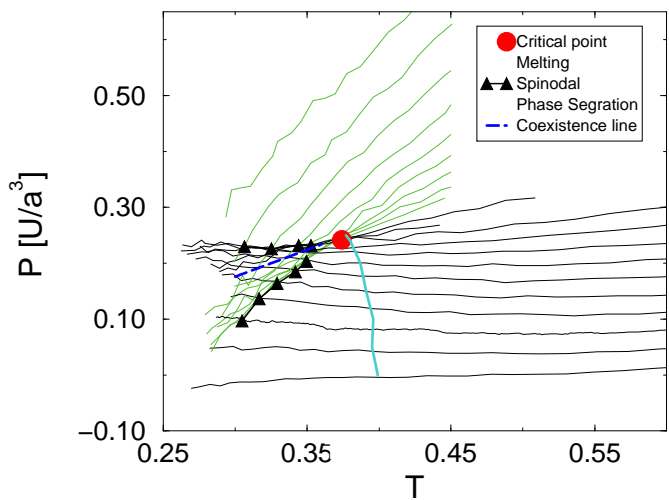}
\caption{The equation of state $P(T,\rho)$ for the ramp potential in the
vicinity of the LL phase transition. Lines indicate P(T) isochores of $21$
densities ($L=15.0,15.2, 15.4,..., 19.0$ from top to bottom). The LL critical
point (closed circle) is located at $P=0.243$, and $T=0.375$, corresponding
to the point at the highest temperature isochore crossing.}
\label{equationofstate}
\end{figure}

\begin{figure}[htb]
\includegraphics[width=8cm,angle=0]{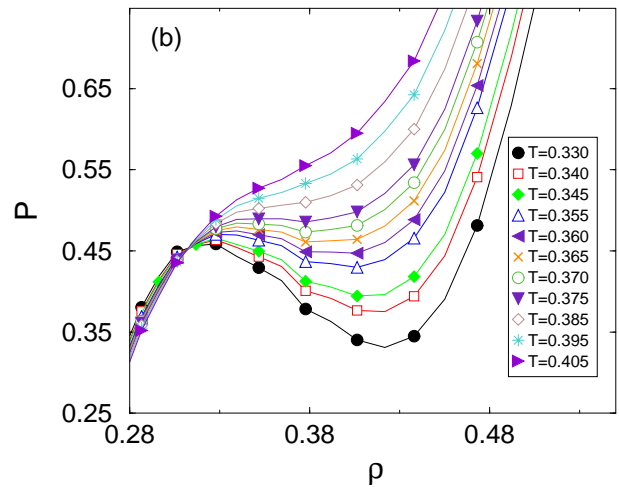}
\caption{The Van der Waals equation of state of the Jagla potential. The
positively sloped LL coexistence line is obtained by the Maxwell rule
construction of the van der Waals equation of state.}
\label{P-V}
\end{figure}

\begin{figure}[htb]
 \centerline{\includegraphics[width=8.5cm,angle=0]{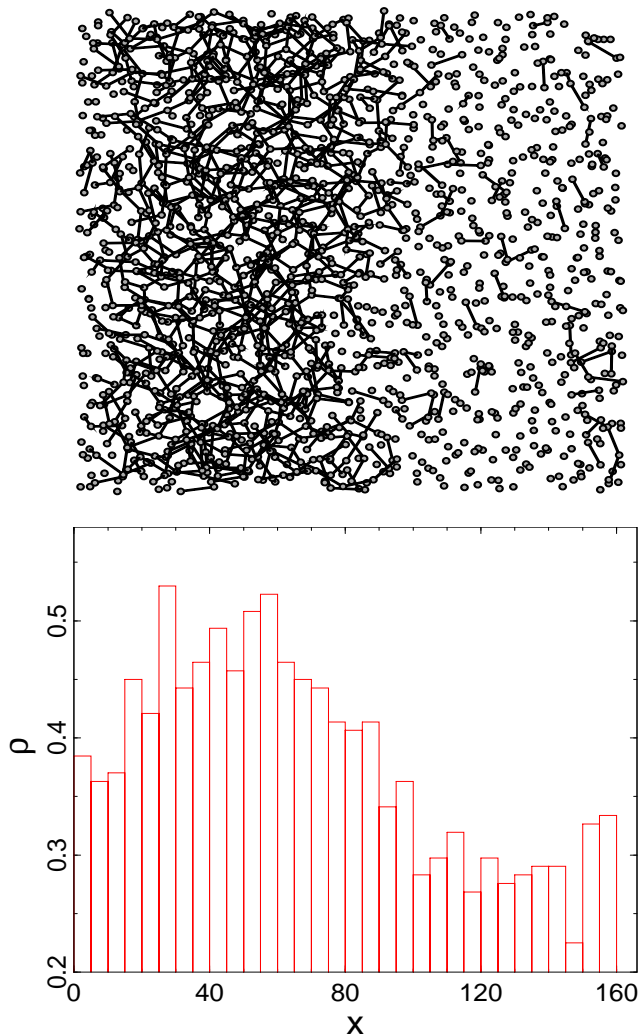}}
 \caption{Phase segregation for system with $L=16.6$ at $T=0.285$. In the top
panel, we present bonds between particles within a distance $d<1.2$,
indicating that the left side is the HDL phase and the right side
corresponding to the LDL phase. The bottom panel is the histogram of
densities along the horizontal direction, with a high density on the left and
lower density on the right. }
\label{phasesegration}
\end{figure}

\section{Results}

The equation of state of the Jagla model [Fig.~\ref{equationofstate}] is
obtained by using constant volume and constant temperature simulations
(NVT-ensemble). The model displays a LL critical point located at
$P_c=0.243\pm0.003$, $T_c=0.375\pm0.002$, and $\rho_c=0.37\pm0.01$. We
determine the LL coexistence line by Maxwell construction on the isotherms
[Fig.~\ref{P-V}]. The coexistence line has a positive slope of $0.96\pm
0.02k_{B}a^{-3}$. According to the Clapeyron equation
\begin{equation}
\frac{dP}{dT}=\frac{\Delta S}{\Delta V},
\end{equation}
the entropy in the HDL phase is lower than entropy in LDL phase. Thus the HDL
phase is more ordered than the LDL phase, which is opposite to the LL
transition found in simulations for water \cite{poole2} and silicon
\cite{Poolenature2001}. The position of the melting line is estimated as the
temperature at which the solid -- liquid first order phase transition occurs
upon gradually heating a system initially consisting of a crystal
configuration obtained by spontaneous crystallization at low temperature. The
equilibrium melting line is about $15\%$ below the upper limit melting
temperatures~\cite{Stilinger_JCP_1985,valeria2005}.
Very recently, it has been suggested that by shrinking the attractive
and repulsive ramp parameters of the Jagla model it is possible to
change the sign of the slope of the LL coexistence line as well as to
move the position of the critical point into the super-cooled region
as in water \cite{Wilding2006}.
 The stability limit is defined by the HDL and low LDL spinodal lines at
 which
\begin{equation}
(\partial P/\partial \rho)_{T}=0.
\end{equation}

We examine the LL phase segregation along each isochore as we cool the system
down in the two-phase region. We did not observe any signs of the phase
segregation at temperatures higher than the spinodal temperature for a given
density. In Fig.~\ref{equationofstate}, the isochores terminate at points
where the phase segregation is clearly visible.  We show the structure of the
$N=1728$ system upon phase segregation, for $L=16.6a$ at $T=0.285$. The
liquid separates into two types of liquids -- HDL and LDL -- with a clear
phase boundary. The density of each phase is consistent with the density at
the coexistent line. We present a clear visualization of the phase boundary
by displaying the bonds between particles within the range $d<1.2a$. In the
LDL phase, there are practically no such particles, while in the HDL phase any
particle has at least one neighbor within this range. This is reflected in
the high peak in the radial distribution function which appears in the HDL
phase~[Fig.~\ref{phasesegration}].

\begin{figure}
 \includegraphics[width=8cm]{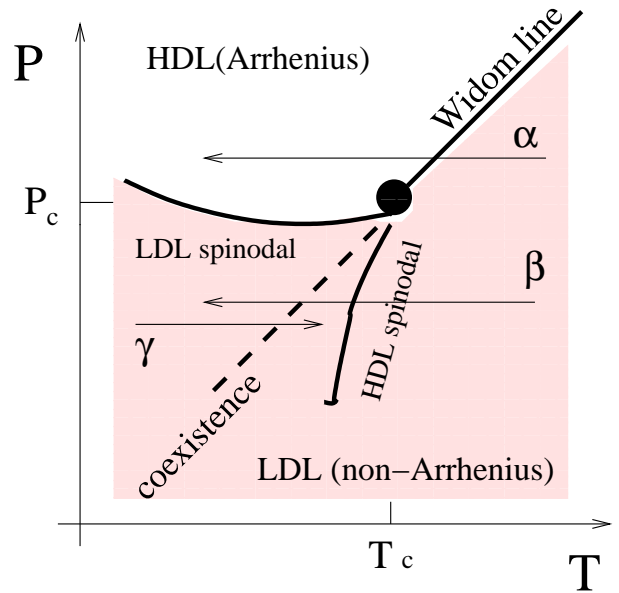}
\caption{ A sketch of the $P-T$ phase diagram for the two-scale Jagla
model. Upon cooling along path $\beta$, the liquid can remain in the LDL
phase (non-Arrhenius) before crossing the LDL stability limit (LDL spinodal
line). Upon heating along path $\gamma$, the liquid may remain in HDL phase
up to the HDL spinodal line. Thus one does not expect any dramatic change in
the dynamic behavior along the path $\beta$ and $\gamma$ in this case. Upon
cooling at constant pressure above the critical point (path $\alpha$), a
dynamic crossover occurs from LDL-like behavior at high-temperature side to
HDL-like behavior at low-temperature side upon crossing the Widom line
\cite{PNAS}.}
\label{phasediagram}
\end{figure}

\begin{figure*}[htb]
\centerline{
  \includegraphics[width=8cm]{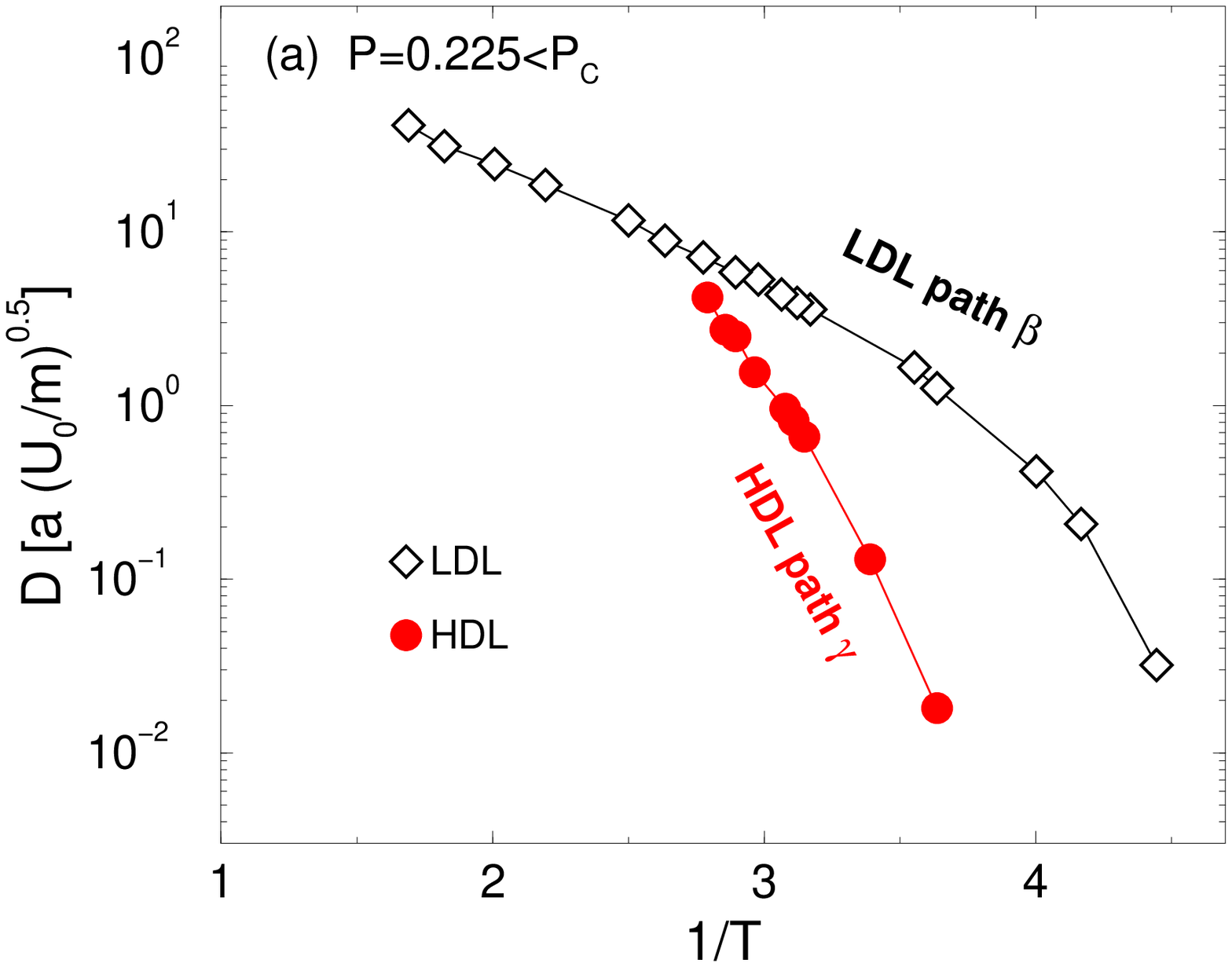}\hfill
  \includegraphics[width=8cm]{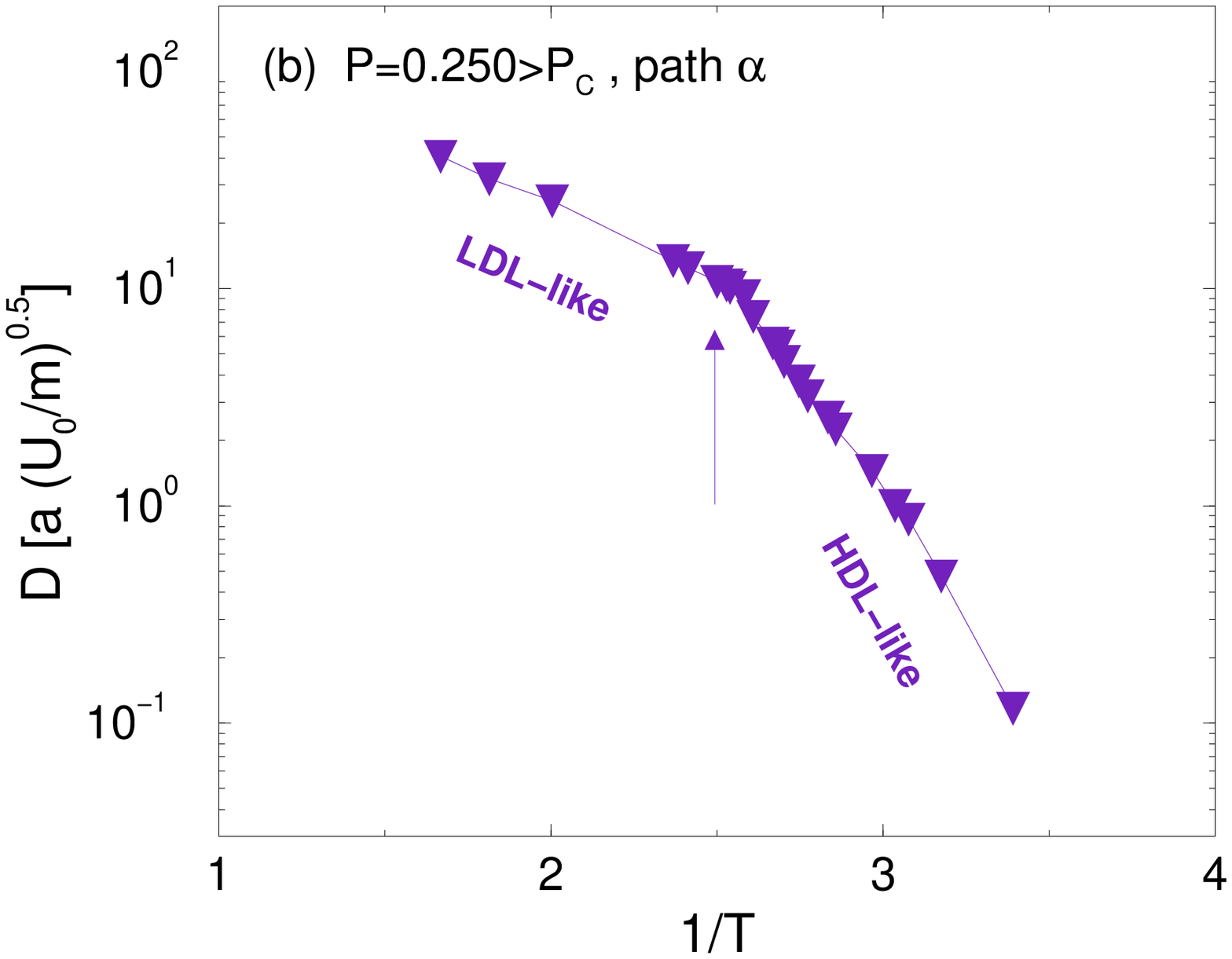}}
\centerline{
  \includegraphics[width=8cm]{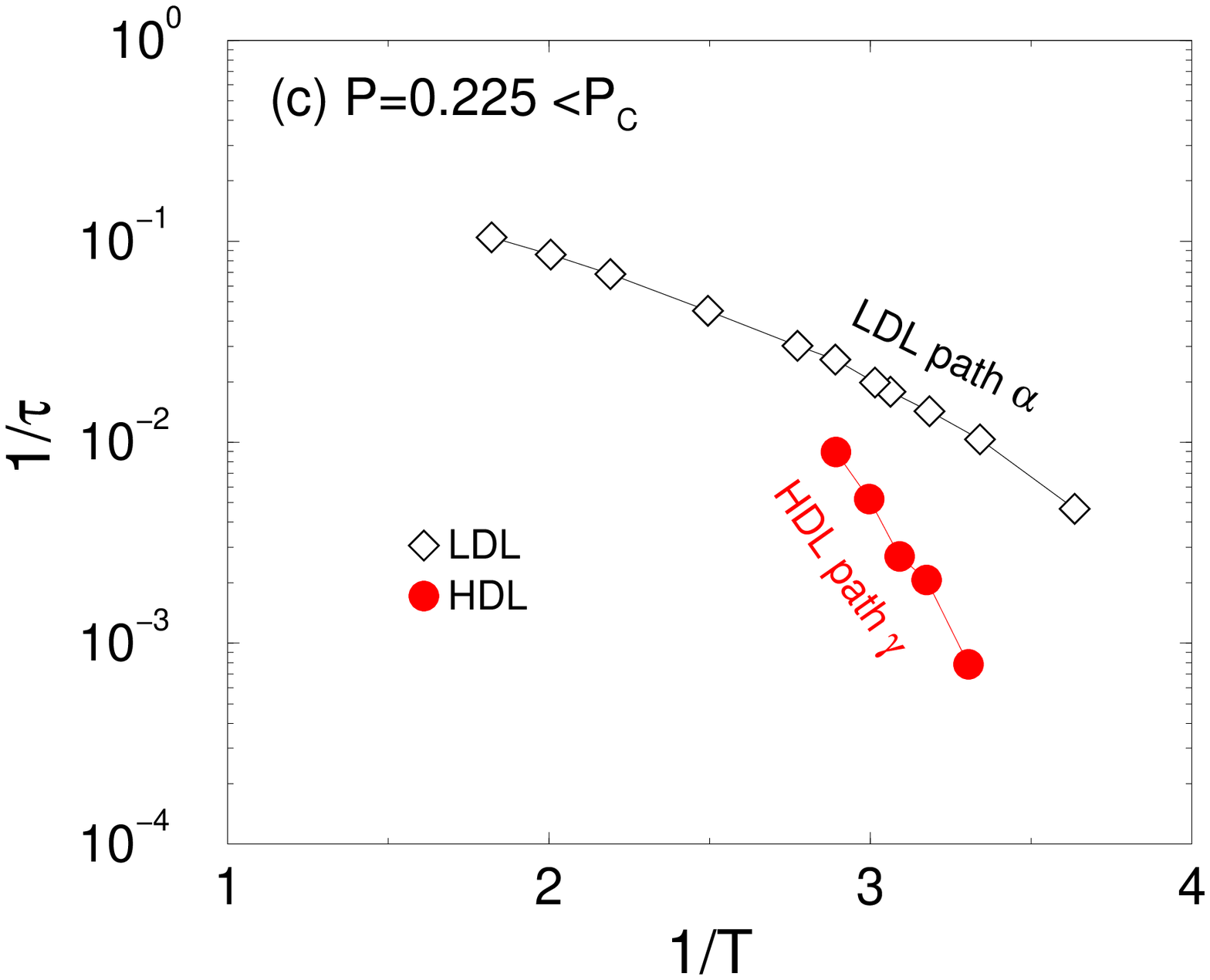}\hfill
  \includegraphics[width=8cm]{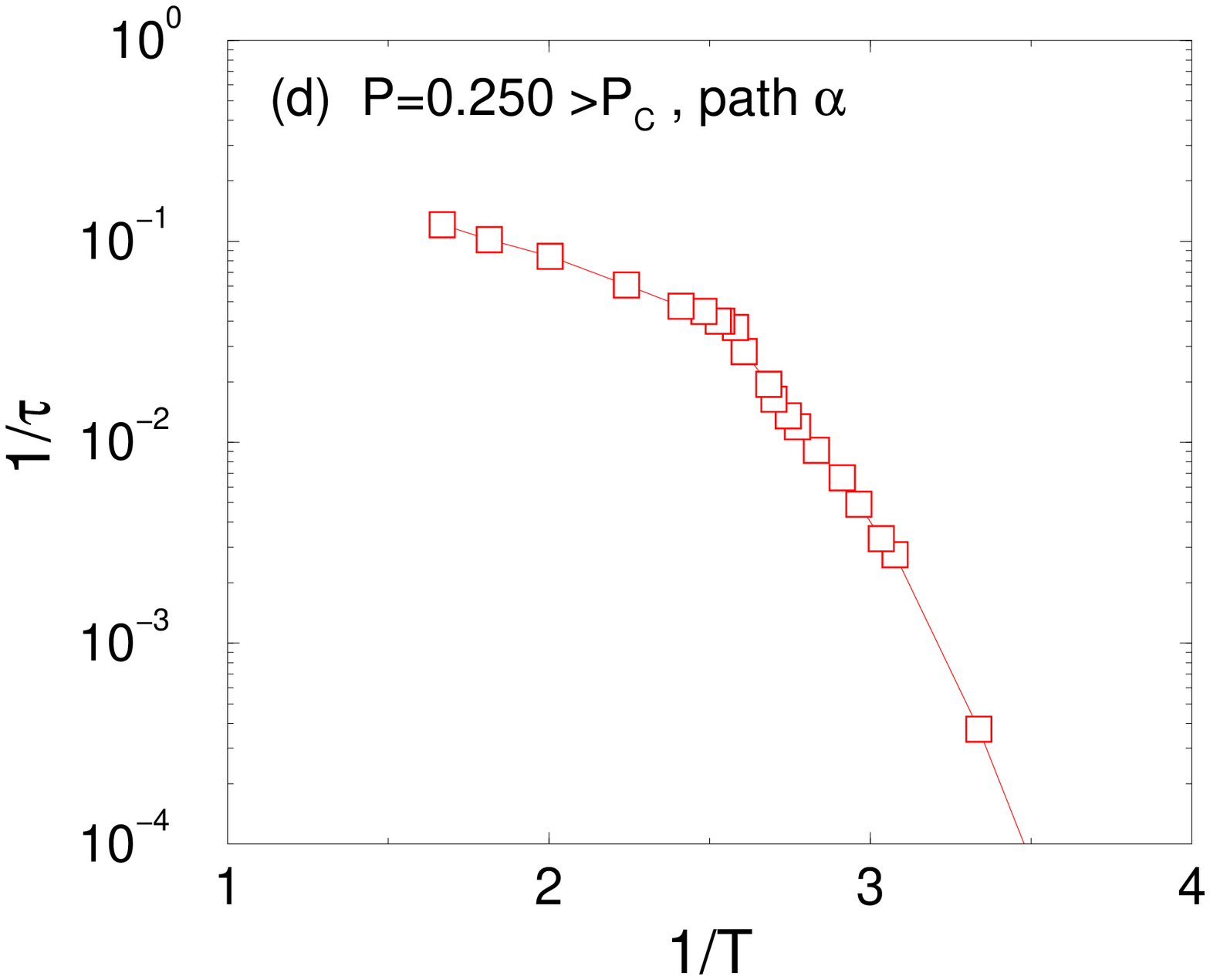}}
\caption{Dynamic behavior for Jagla potential. The $T$-dependence of
diffusivity along constant pressure paths: (a) path $\beta$ with
$P=0.225<P_{c}$ (open diamonds), and path $\gamma$ with $P=0.225<P_{C}$
(closed circles). $D$ in the HDL phase differs from its behavior in the LDL
phase. (b) path $\alpha$ with $0.250>P_{C}$. A dynamic crossover occurs when
the Widom line is crossed upon cooling \cite{PNAS}. (c) Relaxation time
$\tau$ as a function of $1/T$ for $P=0.225$. Similar to $D$, the relaxation
time $\tau$ in the HDL (path $\gamma$) differs from $\tau$ in the LDL (path
$\beta$). (d) Relaxation time $\tau$ along path $\alpha$ with $P=0.250$. A
dynamic crossover, as found in $D$, occurs as the system changes from
``LDL-like'' behavior to ``HDL-like'' behavior.}
\label{D-T-P}
\end{figure*}

We also study the diffusivity $D$ 
\begin{equation}
D\equiv\lim_{t\to \infty}
{\langle\left[{\bf r}(t'+t)-{\bf r}(t')\right]^2 \rangle _{t'} \over
  6t},
\end{equation}
where $\langle\ldots\rangle_{t'}$ denotes an average over all particles and
over all $t'$ along constant pressure paths in three different regions: (i)
$P=0.250$ path $\alpha$ above the critical point in one-phase region
[Fig.~\ref{phasediagram}], (ii) $P=0.225$ path $\beta$
[Fig.~\ref{phasediagram}], below the critical point in the LDL phase, and
(iii) $P=0.225$ path $\gamma$, in the HDL phase [Fig.~\ref{phasediagram}].

Below the LL critical point along path $\gamma$, the diffusivity $D$ exhibits
a temperature behavior different from the behavior in the LDL phase along
path $\beta$. The behavior of the diffusion constant and relaxation time in
the HDL phase follows an approximately Arrhenius law, while in the LDL phase
the slope of the Arrhenius fit increases upon cooling. As we cool along path
$\beta$, we can achieve very low temperatures without phase segregation so
that we can measure the dynamic properties of the LDL in a metastable region
below the coexistence line. 

On the other hand, by heating the HDL phase along path $\gamma$, we can reach
the stability limit of the HDL denoted by a HDL spinodal line. As we cross
the HDL spinodal at constant pressure, the density of abruptly increases and
the diffusivity of the system becomes equal to that of the LDL. In contrast,
along path $\alpha$ in the one phase region above the LL critical point, $D$
exhibits a dynamic crossover from the LDL-like behavior on the high
temperature side of the Widom line to HDL-like behavior on the low
temperature side. This dynamic crossover is an indication of crossing the
Widom line above the LL critical point \cite{PNAS}. 

We also present results for the relaxation time $\tau$ along different paths,
where $\tau$ is defined as the time when the intermediate scattering function
decays to $1/e$ value for a certain $q$ (the first peak of the static
structure factor) \cite{Sciortino}. Along path $\gamma$ with $P=0.225$
[Fig.~\ref{D-T-P}(c)], $\tau$ differs from its behavior in LDL phase along
path $\beta$ [Fig.~\ref{D-T-P}(d)]. Above the LL critical point, along path
$\alpha$ with $P=0.250$, we find that $\tau$ displays a crossover from
LDL-like behavior on the high-temperature side to HDL-like behavior on the
low temperature side, which is similar to the crossover found in $D$.


\section{ Summary}

In summary, we study a simple spherically symmetric two-scale Jagla potential
with both repulsive and attractive parts. The system displays a LL phase
transition well above the equilibrium melting line. The phase segregation shows
direct evidence of the coexistence of two metastable states -- HDL and LDL
phase. The dynamic behaviors in the LDL and HDL are different from each other
below the critical pressure, and change from LDL-like behavior to HDL-like
behavior upon cooling above the critical pressure, suggesting the association
of the changes in dynamic behavior with the LL critical point similar
to the changes observed in water in nan-pores~\cite{chen2005PC}.


We thank C. A. Angell, S.-H Chen, G. Franzese, P. Kumar, J. M. H. Levelt
Sengers, M. Mazza, P. H. Poole, F. Sciortino, S. Sastry, F. W. Starr,
B. Widom, and Z. Yan for helpful discussions and National Science Fundation
grant CHE~0096892 for support. We also thank the Boston University
Computation Center for allocation of CPU time. SVB thanks the Office of
Academic Affairs of the Yeshiva University for funding the high-performance
computer cluster.


\end{document}